# 200 MV RECORD VOLTAGE OF vCM AND LCLS-II-HE CRYOMODULES PRODUCTION START AT FERMILAB*

T. Arkan†, J. Kaluzny, D. Bafia, D. Bice, J. Blowers, A. Cravatta, M. Checchin, B. Giaccone, C. Grimm, B. Hartsell, M. Martinello, T. Nicol, Y. Orlov, S. Posen, Fermilab, Batavia, USA

## Abstract

The Linac Coherent Light Source (LCLS) is an X-ray science facility at SLAC National Accelerator Laboratory. The LCLS-II project (an upgrade to LCLS) is in the commissioning phase; the LCLS-II-HE (High Energy) project is another upgrade to the facility, enabling higher energy operation. An electron beam is accelerated using superconducting radio frequency (SRF) cavities built into cryomodules. It is planned to build 24 1.3 GHz standard cryomodules and one 1.3 GHz single-cavity Buncher Capture Cavity (BCC) cryomodule for the LCLS-II-HE project. Fourteen of these standard cryomodules and the BCC are planned to be assembled and tested at Fermilab. Procurements for standard cryomodule components are nearing completion. The first LCLS-II-HE cryomodule, referred to as the verification cryomodule (vCM) was assembled and tested at Fermilab. Fermilab has completed the assembly of the second cryomodule. This paper presents LCLS-II-HE cryomodule production status at Fermilab, emphasizing the changes done based on the successes, challenges, mitigations, and lessons learned from LCLS-II; validation of the changes with the excellent vCM results.

## INTRODUCTION

LCLS-II-HE cryomodule (CM) production started at Fermilab with the assembly and testing of the verification cryomodule (vCM). Fermilab is responsible for the cryomodule design. The vCM design is the same as the LCLS-II CM; one major difference is that the superconducting radio frequency (SRF) cavities are treated with a new processing protocol for the required performance specifications.

With contributions from Fermilab, Jefferson Lab and SLAC, an R&D effort has been successfully completed to develop the new processing protocol and transfer the technology to industry. Ten fully dressed cavities were fabricated and processed with the newly developed treatment in industry, and successfully tested at Fermilab; performance exceeded the specification with average $Q_0$=3.6e10 and Eacc=25.6 MV/m (specifications are $Q_0$=2.7e10, Eacc=21 MV/m)).

vCM was successfully tested at Fermilab with a 5-month test program and achieved an acceleration voltage of 200 MV in continuous wave mode, corresponding to an average accelerating gradient of 24.1 MV/m, significantly exceeding the specification of 173 MV. The average $Q_0$ (3.0 × $10^{10}$) also exceeded its specification (2.7 × $10^{10}$).

___
* Work supported by U.S. Department of Energy Offices of High Energy Physics and Basic Energy Sciences under Contracts No. DE-AC05-06OR23177 (Fermilab), No. DE-AC02-76F00515 (SLAC),
† email address: arkan@fnal.gov

After quench processing, no field emission was observed up to the maximum gradient of each cavity. At Fermilab, with this world record performance of vCM, we started the production of series cryomodules.

## VCM ASSEMBLY AND TEST

vCM is assembled using eight of the best performing cavities fabricated with the new processing protocol. R&D to develop the new processing protocol, transfer the technology to industry, test and qualify the cavities processed with the new protocol is the first part of the equation to declare success. The second part is to assemble these cavities into the cryomodule and prove that the performance of the cavities can be preserved. vCM was assembled very soon after completing the last LCLS-II CM (keeping the momentum). Based on lessons learned from our successes and from unwanted outcomes of LCLS-II cryomodules production, some infrastructure upgrades were done and validated [1].

vCM cavity string was assembled (Figure 1) in two months which is twice the duration of the LCLS-II production CM string assembly. We had to ensure that the new procedures, tooling, and infrastructure upgrades were fully understood and utilized by the team.

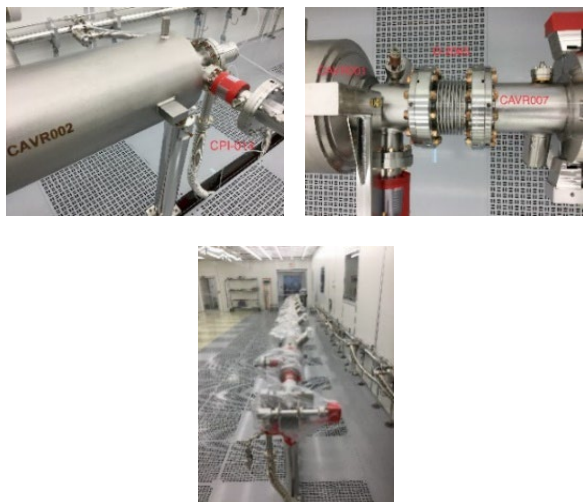

Figure 1: vCM cavity string assembly

A few of the changes done to the cavity string assembly:
- Beamline slow vacuum pumping & nitrogen gas backfill / purge systems upgrades
- Leave beamline under active vacuum during CM assembly, pumping with NEG/ion pump
- Eliminate fundamental power cold coupler (FPC) assembly workstation in the cleanroom and combine the cavity interconnect bellows and cold end

FPC assembly into one workstation. This is mainly done to eliminate additional handling of the cavity with vented beamline and to reduce the opening and closing cycles for the cavity beamline isolation right angle valve.

After the string assembly is completed and rolled out of the cleanroom, cold mass and cryomodule assembly was assembled in four months. We again had to ensure that the new procedures, tooling, and infrastructure upgrades are fully understood and utilized by the team. One of the biggest concerns was assembling the beamline under vacuum. The Fermilab CM assembly team visited JLab to learn and transfer the knowledge for the beamline assembly. We introduced new tooling and procedures to eliminate any risk to the cavity string bellows and unintentional collapse due to beamline vacuum forces. A Failure Mode and Effects Analysis (FMEA) was written to complement the travelers. See Figure 2 for photos of vCM cold mass and CM assembly

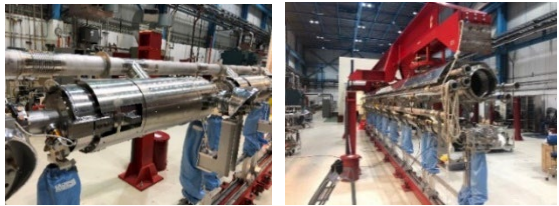

Figure 2: vCM cold mass & CM assembly

During the cold mass assembly at WS2 after the 2-phase circuit welding leak check, a helium vessel leak was found on an SRF cavity. This non-conformance was not experienced before during LCLS-II CM production at Fermilab. Helium vessel leak check is done as a part of the SRF cavity incoming QC. This leak opened during cold test at the vertical test stand. Fortunately, the leak was repaired in situ thanks to Fermilab welding experts. We were able to keep the schedule and did not end up disassembling and moving the cavity string back to the cleanroom to replace the leaky cavity. See Figure 3 for repair. As lessons learned, we revised the production workflow and introduced a new cavity jacket leak check step post vertical test, prior to SRF cavities entering the cleanroom for string assembly.

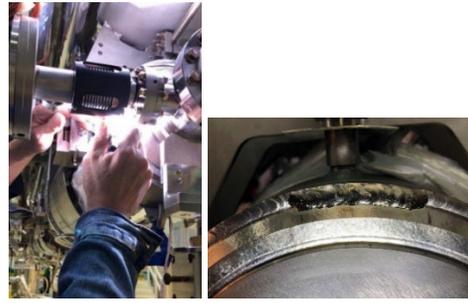

Figure 3: Helium vessel leak repair in situ on a dressed cavity during cold mass assembly

The assembly of vCM was completed at the end of March 2021 and it was transported to the Fermilab CM Test Stand (CMTS). We used the proven, established onsite CM transport procedures and experienced team to transport the CM from the assembly floor to CMTS as seen in Figure 4.

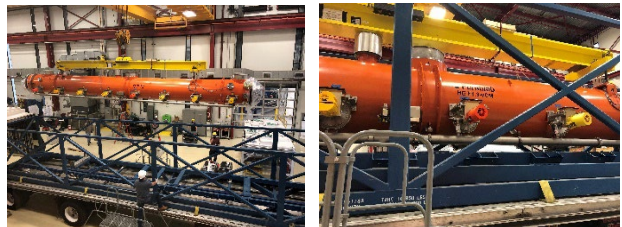

Figure 4: CM transport to CMTS

vCM is tested at CMTS [4] (See Figure 5).

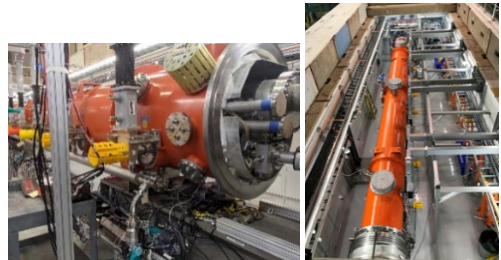

Figure 5: vCM at CMTS

Cavity performance results look excellent. This is a world record CW CM. Gradient and $Q_0$ in all eight cavities exceed the LCLS-II-HE specification [2] and are well above average compared to LCLS-II production. The CM is also almost field emission free: only a very minimal level of x-rays (~1 mR/hr) were detected while powering one cavity at the operating gradient and this was processed away during the unit test. We also applied plasma processing to this cryomodule even though it was field emission free. Post plasma processing results showed that no new field emission was introduced to the CM. One benefit observed after plasma processing was the elimination of multipacting quenches [3]. These excellent results prove that all the work done to prepare for LCLS-II-HE CM assembly, keeping the momentum from LCLS-II and applying the lessons learned, was an effective strategy. vCM has successfully shipped and delivered to SLAC in February 2022. See Figure 6.

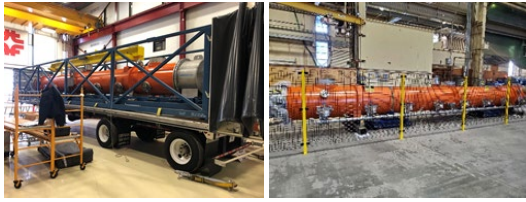

Figure 6: vCM prepped for shipping at Fermilab and stored at SLAC

## PRODUCTION CRYOMODULES

Procurement of parts for series cryomodules production is nearing completion at Fermilab, see Figure 7. SRF cavities and fundamental power coupler procurements for the project will continue at SLAC and JLab throughout 2023.

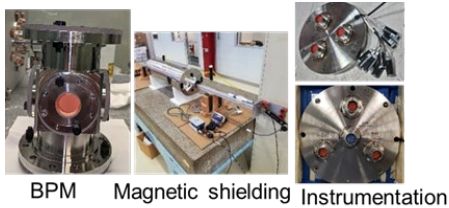

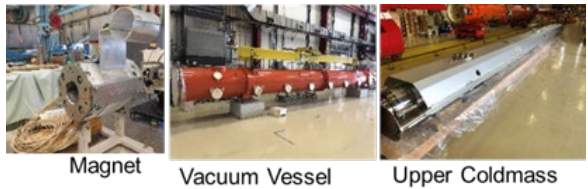

Figure 7: Cryomodule parts received, in quality control at Fermilab

The first article CM has been fully assembled and tested at Fermilab. Excellent results from first article re-affirms production readiness at Fermilab, this CM will be shipped to SLAC in September 2022, see figure 8.

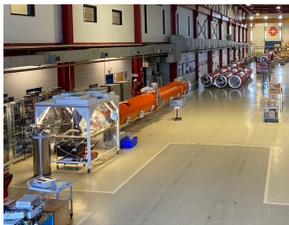

Figure 8: First article CM being prepared for shipping to SLAC

We continue to receive dressed cavities from the cavity vendor and test/qualify them individually with cold vertical test before string assembly. The second production CM assembly is complete and ready for testing at CMTS. Fermilab plans to produce and deliver all series production CMs by the end of 2025.

## BUNCHER CAVITY CRYOMODULE

The LCLS-II-HE project scope includes the construction of a 100 MeV low emittance injector (LEI). The LEI includes two SRF cavities: an electron gun cavity and a 1.3 GHz, 9-cell buncher cavity. The latter is housed in the Buncher Cavity Cryomodule (BCC) that includes a solenoid magnet package. The 1.3-GHz dressed cavity is to be of the same design as that used in the standard eight-cavity 1.3 GHz CMs including the same FPC and tuner. The BCC CM will have a BPM identical to the standard CM and will have a solenoid magnet with corrector and quad trim windings. See Figure 9. Fermilab will design, build, test and deliver the BCC to SLAC in early 2026.

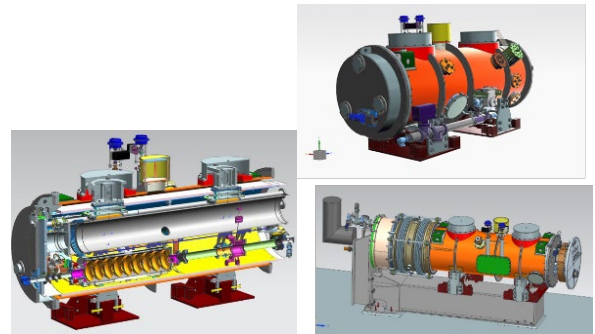

Figure 9: BCC design in progress

## CONCLUSION

LCLS-II CM production at Fermilab has brought technical readiness and important lessons learned for LCLS-II-HE. We have developed a strong team. We have proven that our team of people worked well together internally and externally and produced excellent results. The Fermilab team has demonstrated the capability of producing world record performance cryomodules on schedule and on budget. The push for even higher performance for LCLS-II-HE is progressing well with the vCM successful assembly and excellent test results. The Fermilab infrastructure and team are fully committed to complete its scope for the LCLS-II-HE.